\documentclass[iop,twocolappendix,numberedappendix]{emulateapj}
\usepackage{epsfig}
\usepackage{graphicx}
\usepackage{subfigure}
\usepackage{multirow}
\usepackage{amssymb}
\usepackage{natbib}
\usepackage{rotating}
\usepackage[bookmarks=false, colorlinks=true, citecolor=blue, backref=true]{hyperref}
\usepackage{apjfonts}

\shorttitle{ALMA Observations of NGC~34}
\shortauthors{C.K. Xu et al.}

\newcommand{\lsim}{\, \lower2truept\hbox{${< \atop\hbox{\raise4truept\hbox{$\sim$}}}$}\,}
\newcommand{\gsim}{\, \lower2truept\hbox{${> \atop\hbox{\raise4truept\hbox{$\sim$}}}$}\,}

\begin{document}

\slugcomment{{\bf Draft 11}; \today}

\title{ALMA Observations of Warm Molecular Gas and Cold Dust in NGC~34}
\thanks{The National Radio Astronomy Observatory is a facility of the National Science Foundation operated under cooperative agreement by Associated Universities, Inc.}
\author{
C.~K.~Xu\altaffilmark{1}, 
C.~Cao\altaffilmark{2,3,1}, 
N.~Lu\altaffilmark{1},
Y.~Gao\altaffilmark{4},
P.~van~der~Werf\altaffilmark{5},
A.~S.~Evans\altaffilmark{6,7},
J.~M.~Mazzarella\altaffilmark{1},
J.~Chu\altaffilmark{8},
S.~Haan\altaffilmark{9},
T.~Diaz-Santos\altaffilmark{1},
R.~Meijerink\altaffilmark{5},
Y.-H.~Zhao\altaffilmark{1,4},
P.~Appleton\altaffilmark{1},
L.~Armus\altaffilmark{1},
V.~Charmandaris\altaffilmark{10,11,12},
S.~Lord\altaffilmark{1},
E.~J.~Murphy\altaffilmark{1},
D.~B.~Sanders\altaffilmark{8},
B.~Schulz\altaffilmark{1},
S.~Stierwalt\altaffilmark{6,13}
}
\altaffiltext{1}{Infrared Processing and Analysis Center, MS 100-22, California Institute of Technology, Pasadena, CA 91125}
\altaffiltext{2}{School of Space Science and Physics, Shandong University at Weihai, Weihai, Shandong 264209, China, caochen@sdu.edu.cn}
\altaffiltext{3}{Shandong Provincial Key Laboratory of Optical Astronomy and Solar-Terrestrial Environment, Weihai, Shandong 264209, China}
\altaffiltext{4}{Purple Mountain Observatory, Chinese Academy of Sciences, 2 West Beijing Road, Nanjing 210008, China}
\altaffiltext{5}{Leiden Observatory, Leiden University, P.O. Box 9513, NL-2300 RA Leiden, Netherlands}
\altaffiltext{6}{Department of Astronomy, University of Virginia, P.O. Box 400325, Charlottesville, VA 22904, USA}
\altaffiltext{7}{National Radio Astronomy Observatory, Charlottesville, VA 22904, USA}
\altaffiltext{8}{Institute for Astronomy, University of Hawaii, 2680 Woodlawn Dr., Honolulu HI 96816, USA}
\altaffiltext{9}{CSIRO Astronomy and Space Science, ATNF, PO Box 76, Epping 1710, Australia}
\altaffiltext{10}{Department of Physics, University of Crete, GR-71003, Heraklion, Greece}
\altaffiltext{11}{Institute for Astronomy, Astrophysics, Space Applications \& Remote Sensing, National Observatory of Athens, GR-15236, Penteli, Greece}
\altaffiltext{12}{Chercheur Associ\'e, Observatoire de Paris, F-75014,  Paris, France}
\altaffiltext{13}{Spitzer Science Center, California Institute of Technology, 1200 E. California Blvd., Pasadena, CA 91125}

\begin{abstract}
We present ALMA Cycle-0 observations of the CO~(6-5) line emission
(rest-frame frequency = 691.473 GHz) and of the 435$\mu m$ dust
continuum emission in the nuclear region of NGC~34, a local luminous
infrared galaxy (LIRG) at a distance of 84~Mpc ($\rm 1\arcsec = 407\;
pc$) which contains a Seyfert~2 active galactic nucleus (AGN) and a
nuclear starburst.  The CO emission is well resolved by the ALMA beam
($\rm 0\farcs26\times 0\farcs23$), with an integrated flux of $\rm
f_{CO~(6-5)} = 1004\; (\pm 151) \; Jy\; km\; s^{-1}$. Both the
morphology and kinematics of the CO~(6-5) emission are rather regular,
consistent with a compact rotating disk with a size of 200 pc. 
A significant emission feature
is detected on the red-shifted wing of the line profile at the
frequency of the $\rm H^{13}CN\; (8-7)$ line, with an integrated flux
of $\rm 17.7 \pm 2.1 (random) \pm 2.7 (sysmatic)\; Jy\;km\;
s^{-1}$. However, it cannot be ruled out that the feature is due to an
outflow of warm dense gas with a mean velocity of $\rm 400\; km\;
s^{-1}$.  The continuum is resolved into an elongated configuration,
and the observed flux corresponds to a dust mass of $\rm M_{dust} = 10^{6.97\pm 0.13}\;
M_\sun$. An unresolved central core ($\rm radius \simeq 50\; pc$)
contributes $28\%$ of the continuum flux and $19\%$ of the CO~(6-5)
flux, consistent with insignificant contributions of the AGN to both
emissions.  Both the CO~(6-5) and continuum spatial distributions
suggest a very high gas column density ($\rm \gsim 10^4\; M_\sun\;
pc^{-2}$) in the nuclear region at $\rm radius \lsim 100\; pc$.
\end{abstract}


\vskip3truecm

\section{Introduction}

\begin{deluxetable*}{cccccccc}
\tabletypesize{\normalsize}
\setlength{\tabcolsep}{0.05in} 
\tablecaption{ALMA Observations \label{tbl:obs}}
\tablehead{
 {SB}  & Date & {Time (UTC)}  & {Config}  & {$\rm N_{ant}$} & {$\rm l_{max}$} & {$\rm t_{int}$} & {$\rm T_{sys}$}\\
       & (yyyy/mm/dd) &  &   &  & (m) & {(min)} & {(K)}\\
{(1)} & {(2)} & {(3)} & {(4)} & {(5)} & {(6)} & {(7)}  & {(8)}
}
\startdata
X40e374\_Xba3 & 2012/05/20 & 09:17:15 -- 10:35:34 & E & 16 & 375&24.7 &850 \\
X40e374\_Xd36 & 2012/05/20 & 10:48:14 -- 12:06:33 & E & 16 & 375&24.7 &653 \\
X40e374\_Xec9 & 2012/05/20 & 12:22:31 -- 13:07:05 & E & 16 & 375& 9.9 &654 \\
X41065b\_X334 & 2012/05/21 & 09:47:04 -- 11:06:00 & E & 16 & 375&24.7 &634 \\
X41065b\_X4c7 & 2012/05/21 & 11:20:00 -- 10:41:09 & E & 16 & 375&24.7 &528 \\
X4afc39\_X43b & 2012/08/25 & 03:30:38 -- 05:01:41 &E\&C&27 & 402&24.7 &1058
\enddata
\tablecomments{Column (1) -- schedule-block number; (2) \& (3) -- observation
date and time; (4) -- configuration; (5) -- number of antennae;
(6) -- maximum baseline length; (7) -- on-target integration time;
(8) -- median $\rm T_{sys}$.
}
\end{deluxetable*}

Luminous infrared galaxies (LIRGs: $\rm L_{IR} [8 \hbox{--} 1000\mu m]
> 10^{11} \; L_\sun$), including ultra-luminous infrared galaxies
(ULIRGs: $\rm L_{IR} > 10^{12}\; L_\sun$), have a space density exceeding
that of optically selected starburst and Seyfert galaxies at
comparable bolometric luminosity \citep{Soifer1987}.  They are an
ensemble of single galaxies, galaxy pairs, interacting systems and
advanced mergers \citep{Sanders1996, Wang2006}. Most (U)LIRGs of $\rm
L_{IR} \gsim 10^{11.5}\; L_\sun$ are advanced mergers (including merger
remnants), harboring extreme starbursts (star formation rate (SFR)
$\rm \gsim 50\; M_\sun\; yr^{-1}$) and powerful AGNs \citep{Kim2002}.
\citet{Toomre1978} was the first to suggest that merging can
transform spirals into ellipticals, a theory that has been borne out
observationally \citep{Schweizer1982, Genzel2001}. Strong outflows
of neutral and ionized gas were widely detected among (U)LIRGs
\citep{Armus1990, 
Heckman1990, Rupke2005}. Recently, massive molecular 
gas outflows have been found in (U)LIRGs with powerful AGNs \citep{Fischer2010,
Feruglio2010, Sturm2011, Aalto2012b, 
Veilleux2013, Feruglio2013, Combes2013, Cicone2014}. 
Hence, feedback from merger induced extreme starbursts and AGNs is
the most popular mechanism for explaining the star formation quenching in
massive galaxies that lead to the formation of red sequence galaxies
\citep{Faber2007, Hopkins2008a}, and (U)LIRGs are the best
local laboratories for studying these processes.  However, 
due to the enormous dust obscuration in (U)LIRGs 
\citep{Sanders1996}, it is very
difficult to study them using
high angular resolution optical/NIR instruments.
Observations using
space FIR/sub-mm observatories, such as Spitzer 
\citep{Werner2004} and Herschel \citep{Pilbratt2010},
 can penetrate the dust obscuration. However, their angular resolutions
($\gsim$ a few arcsecs) are not sufficient to resolve the 
nuclei in most (U)LIRGs. The Atacama Large Millimeter Array 
(ALMA; \citealt{Wootten2009}) 
is changing the situation rapidly.
Once completed, sub-mm/mm observations using the ALMA full
array (66 antennae) will detect both the line emission of gas and the continuum
emission of dust (heated either by starburst or AGN, or both) with an
angular resolution of $\lsim 0.1''$, revealing interplays between
gas, starbursts and AGNs in (U)LIRG nuclei down to linear
scales of $\sim$ 10 pc in the nearest systems.

In this paper, we report ALMA Cycle-0 observations (utilizing up to 27 antennae)
of the CO~(6-5) line (rest-frame frequency = 691.473 GHz) 
emission and of 435$\mu m$ dust continuum emission in the
nuclear region of NGC~34, a local LIRG ($\rm L_{IR} = 10^{11.49}\; L_\sun$,
\citealt{Armus2009}) containing a Sy2 AGN and a nuclear
starburst (see \citealt{Schweizer2007} for an overview). 
With an excitation temperature of $\rm T_{ex} = 116.2\; K$ and
a critical density of $\rm n_{crit} = 2.9\times 10^{5}\; cm^{-3}$
\citep{Carilli2013}, the
CO~(6-5) line probes the warm and dense gas that is much more
closely related to the star formation activities than the cold and
lower density gas probed by low J CO lines (e.g. CO~(1-0) and
CO~(2-1)) commonly studied in the literature. 
Among the complete sample of 202 LIRGs of the Great
Observatories All-sky LIRG Survey (GOALS; \citealt{Armus2009}), which
were selected from the IRAS Revised Bright Galaxy Sample (RBGS;
\citealt{Sanders2003}), NGC~34 (also known as NGC~17, Mrk~938, 
and IRAS~F00085-1223) was chosen for early ALMA observations
because of the following
features: (1) Among LIRGs that have the CO~(6-5) line flux
$\rm f_{CO~(6-5)} \geq 1000\; Jy\; km\; s^{-1}$ observed in the the
Herschel SPIRE Fourier Transform Spectrometer (FTS) survey of GOALS galaxies
(angular resolution: $\sim 30''$; 
van~der~Werf et al., in preparation; Lu et al., in preparation),
it is one of the closest with a distance of $\rm D = 84$ Mpc 
($\rm 1'' = 407 pc$). This enables high signal-to-noise ratio
observations of warm gas structures with 
the best linear resolution for a given
angular resolution.  (2) With a declination of $-12^\circ$,
NGC~34 transits at $\sim 11^\circ$ from the
zenith, therefore its Band 9 observations are affected by minimal
atmospheric absorption.  (3) The Keck-MIRLIN images of \citet{Gorjian2004}
detected $\gsim 50\%$ of the 12$\mu m$ IRAS flux in the central $\sim
1$ kpc region of NGC~34, indicating strong nuclear star formation and/or
AGN activities.

Early observations of the CO~(6-5) emission in nearby starburst galaxies
NGC~253, IC~342 and M~82 \citep{Harris1991} showed that the molecular
gas in starbursts is warmer than in normal disk galaxies.
\citet{Papadopoulos2012} carried out an extensive survey for higher J
CO lines, including the CO~(6-5) line, for LIRGs using JCMT and found
many of them have the global CO spectral line energy distribution
(SLED) dominated by a very warm ($\rm T \sim 100 K$) and dense ($\rm n
\geq 10^4\; cm^{−3}$) gas phase. The 
CO~(6-5) map of the
central region of Arp~220 obtained by \citet{Matsushita2009} using the SMA 
has an angular resolution of $\sim 1\arcsec$ ($\rm \sim 400\;
pc$), revealing two warm gas components 
associated with the two nuclei (separation $\rm \sim 400\;
pc$), but the data were unable to resolve the individual components.
The SMA observations of the CO~(6-5) line emission of VV~114
\citep{Sliwa2013} has a relatively coarse resolution of $\rm 2\farcs5$
($\rm \gsim 1\; kpc$). The new ALMA observations reported here, with an angular
resolution of $\rm \sim 0\farcs25$, resolved the warm and dense gas in
the nuclear region of a LIRG with a linear resolution of $\sim 100$ pc
for the first time. 
We aim to answer the following
questions with these observations: (1) How is the warm dense molecular
gas distributed in the inner most region of NGC~34?  (2) How is the
gas related to the starburst and the AGN, respectively?  (3) Is the
dust emission dominated by an extended component, or by an unresolved
compact component?  (4) Is there evidence for a molecular outflow?
Comparing with the extensive literature on NGC~34 (a late stage
major-merger, \citealt{Mazzarella1993, Schweizer2007}) and with
high-resolution simulations \citep{Hopkins2013a}, our results shed new
light on how various activities and their feedback set the stage for
a major-merger remnant to become a red-and-dead elliptical galaxy
\citep{Zubovas2012}.

\section{Observations}
Observations of the CO~(6-5) line emission and 435~$\mu m$ dust continuum
emission in the nuclear region of NGC~34 were carried out using the
Band 9 receivers of ALMA in the TDM mode (velocity resolution: 6.8
km~sec$^{-1}$). The four basebands (i.e. ``Spectral Windows'',
hereafter SPWs) are centered at the sky frequencies of 679.801,
678.039, 676.271 and 674.339 GHz, respectively, each with a bandwidth
of $\sim 2$ GHz. Observations were carried out in both extended (E)
and extend \& compact (E\&C) configurations (Table~\ref{tbl:obs}).
The total on-target integration time was 2.25 hours. 
During the observations, phase and gain variations
were monitored using QSOs 2348-165 and 3C~454.3. Observations of
the asteroid Pallas were made for flux calibration. 
The error in the flux calibration was estimated to be 15\%.

\begin{figure}[!htb]
\epsscale{1.5}
\vskip-0.5truecm
\plotone{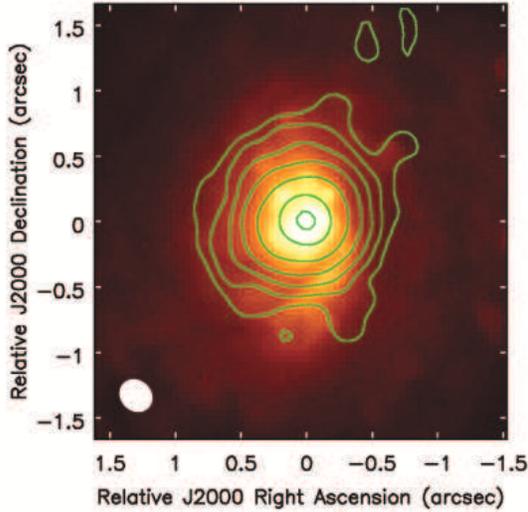}
\vskip-6.5truecm
\caption{Contours of the integrated
CO~(6-5) line emission
of NGC~34 on the HST V-band image
\citep{Malken1998}. The contour levels are 
[1, 2, 4, 8, 16, 32, 50]$\rm \times 3.6\; Jy\; beam^{-1} \; km\; s^{-1}$.
The white ellipse shows the synthesized beam size (FWHM =
$\rm 0\farcs26\times 0\farcs23$, $\rm P.A.=46^\circ.3$). 
The map center is at $\rm R.A. (J2000) = 00^{h}11^{m}6^{s}.537$ and
$\rm Dec (J2000) = -12^\circ 06\arcmin 27\farcs49$.}
\label{fig:integrated}
\end{figure}

\begin{figure}[!htb]
\epsscale{1.25}
\vskip-0.5truecm
\plotone{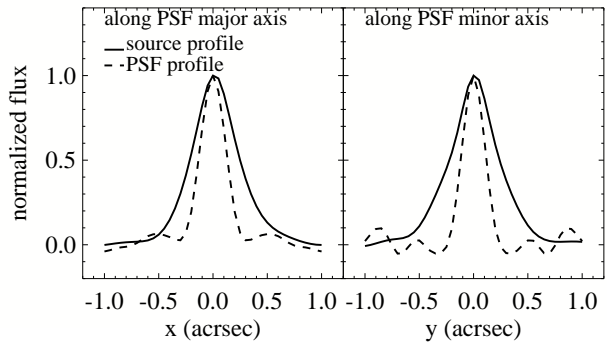}
\caption{Comparisons of profiles of the
integrated CO~(6-5) line emission and the synthesized beam (FWHM =$\rm 0\farcs26\times 0\farcs23$, $\rm P.A.=46^\circ.3$).
}
\label{fig:psf_line}
\end{figure}
\begin{figure}[!htb]
\epsscale{1.0}
\plotone{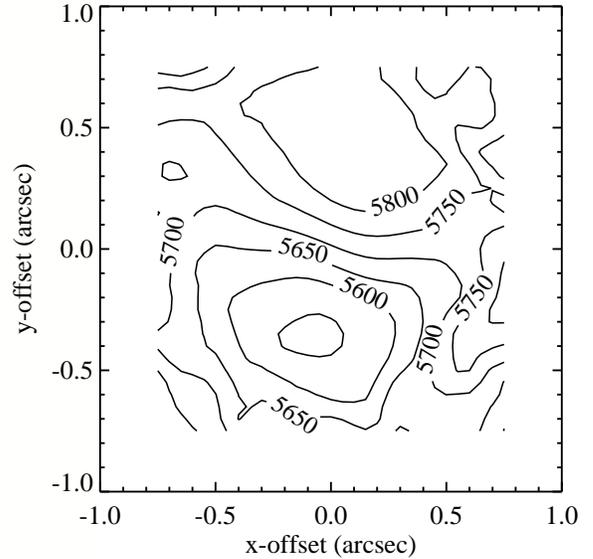}
\vspace{0.3truecm}
\caption{Contour map of the first moment. The LSR (radio)
velocity contour levels are in the units of $\rm km\; s^{-1}$.}
\label{fig:1stmom}
\end{figure}

\begin{figure*}[!htb]
\epsscale{1.0}
\plotone{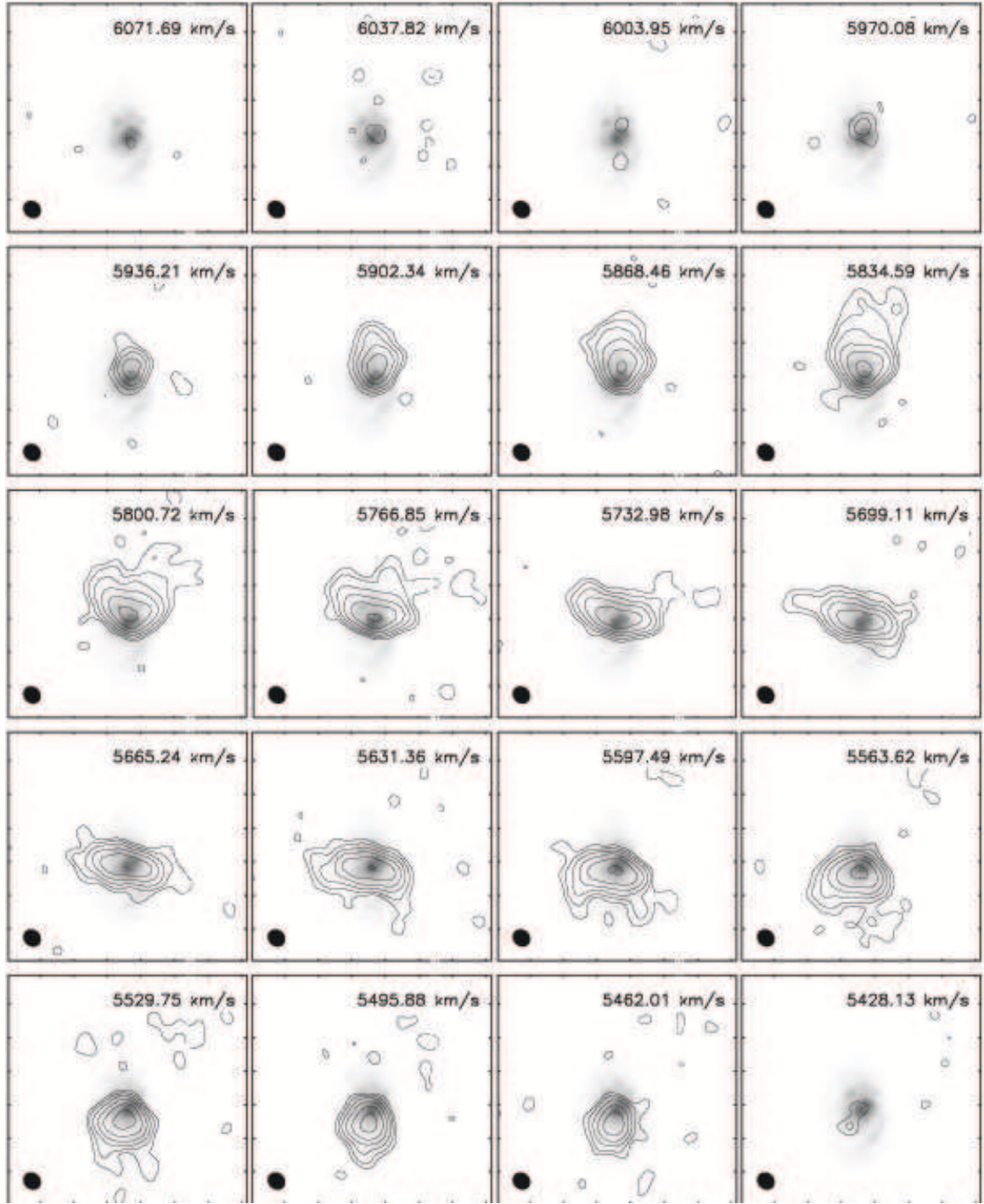}
\caption{CO(6-5) line emission contours of the channel maps 
(velocity channel width = $\rm 34\; km\; s^{-1}$), 
overlaid on the HST V-band image. The contour levels are
$\rm 16.5\; mJy\; beam^{-1} \times$ [1, 2, 4, 8, 16, 24]. All maps have the same
size of $\rm 3''.4\times 3''.4$ and the tick marks are separated by 0\farcs5.
In each channel, the central LSR (radio) velocity is given.
}
\label{fig:channel}
\end{figure*}

\begin{figure}[!htb]
\epsscale{1.0}
\plotone{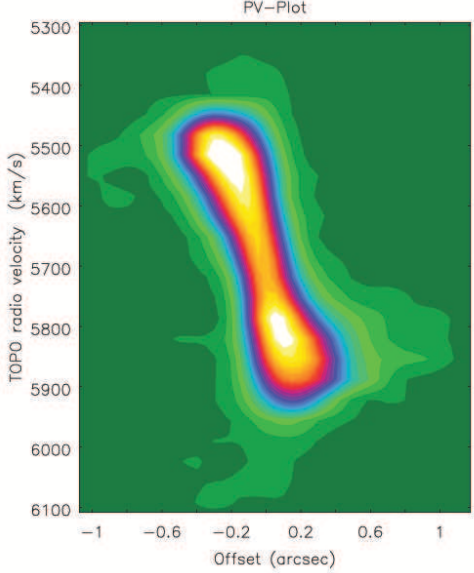}
\caption{P-V plot of the CO~(6-5) line emission over a stripe 
(with the width of $0\farcs25$ and $\rm P.A. = 345^\circ$)
passing through the central core.}
\label{fig:pvplot}
\end{figure}

\begin{figure}[!htb]
\epsscale{1.2}
\plotone{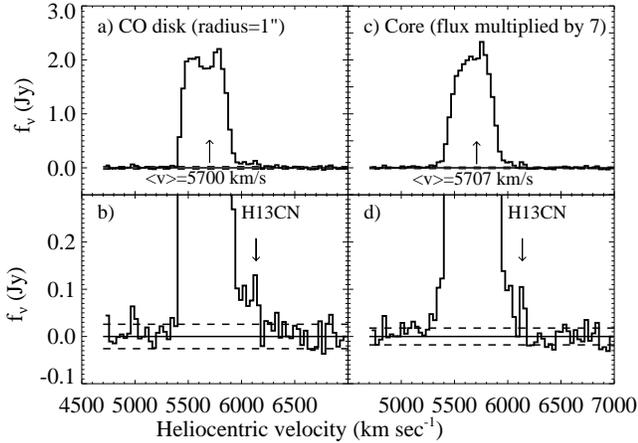}
\caption{
{\bf Panel a:} Spectrum of the CO~(6-5) line emission
in radio heliocentric velocity domain, measured in the channel maps
with an aperture of $\rm radius=1''=407\; pc$. The dashed lines mark the
1-$\sigma$ rms boundaries, which are $\rm \pm 26\; mJy\; 
\times 34\; km\; s^{-1}$.
{\bf Panel b:} Zoom-in plot of the bottom part of {\it Panel a}. 
Again the dashed lines mark the 1-$\sigma$ rms boundaries. The arrow
marks the location of the $\rm H^{13}CN$~(8-7) line with the velocity
$\rm v = 5745\; km\; s^{-1}$.
{\bf Panel c:} Spectrum of the CO~(6-5) line emission of the central
core ($\rm radius=0\farcs125=52\; pc$). The flux is scaled up by a factor of 7. 
{\bf Panel d:} Zoom-in plot of the bottom part of {\it Panel c}. 
The arrow
marks the location of the $\rm H^{13}CN$~(8-7) line with the velocity
$\rm v = 5745\; km\; s^{-1}$.
}
\label{fig:vdis}
\end{figure}

\begin{figure}[!htb]
\epsscale{0.7}
\hspace{-1truecm}\plotone{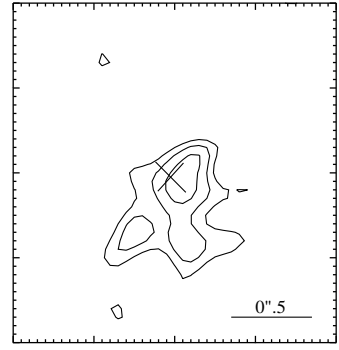}
\caption{Contour map of the red-shifted feature
(integrated over channels in velocity range of
 $\rm 6000\; km\; s^{-1} < v < 6200\; km\; s^{-1}$).
The contour levels are  $\rm \sigma \times$ [3, 4, 5],
and $\rm \sigma = 0.46\; Jy\; beam^{-1} km\; s^{-1}$ (rms).
The central cross marks the position of the CO~(6-5) line emission peak,
and lengths of the two axes of the cross represent the
the major and minor axes of the ALMA beam. The map is 
2\arcsec$\times$2\arcsec in size.
}
\label{fig:h13cn}
\end{figure}

\begin{figure}[!htb]
\epsscale{0.9}
\plotone{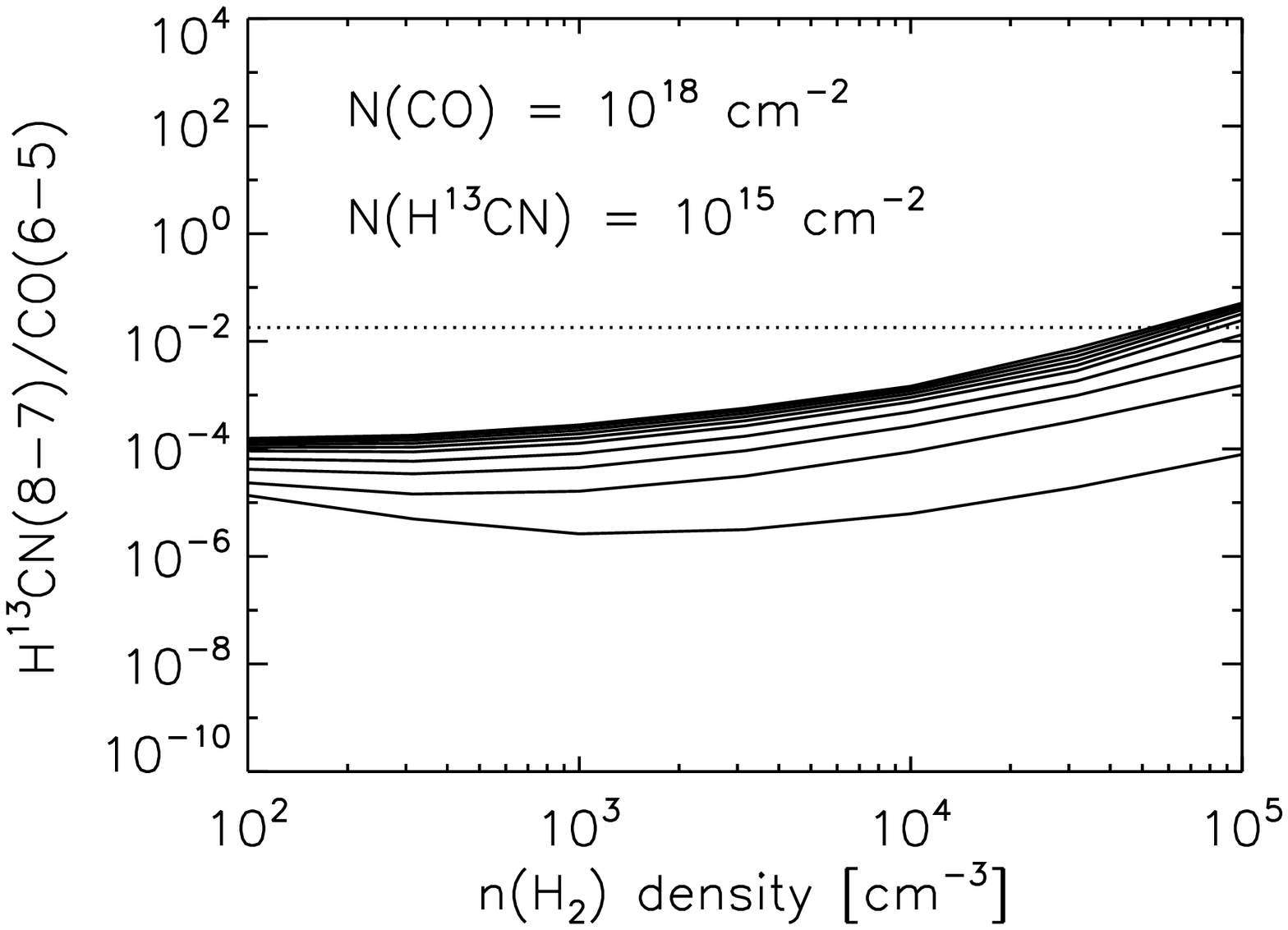}
\vspace{-0.5truecm}
\plotone{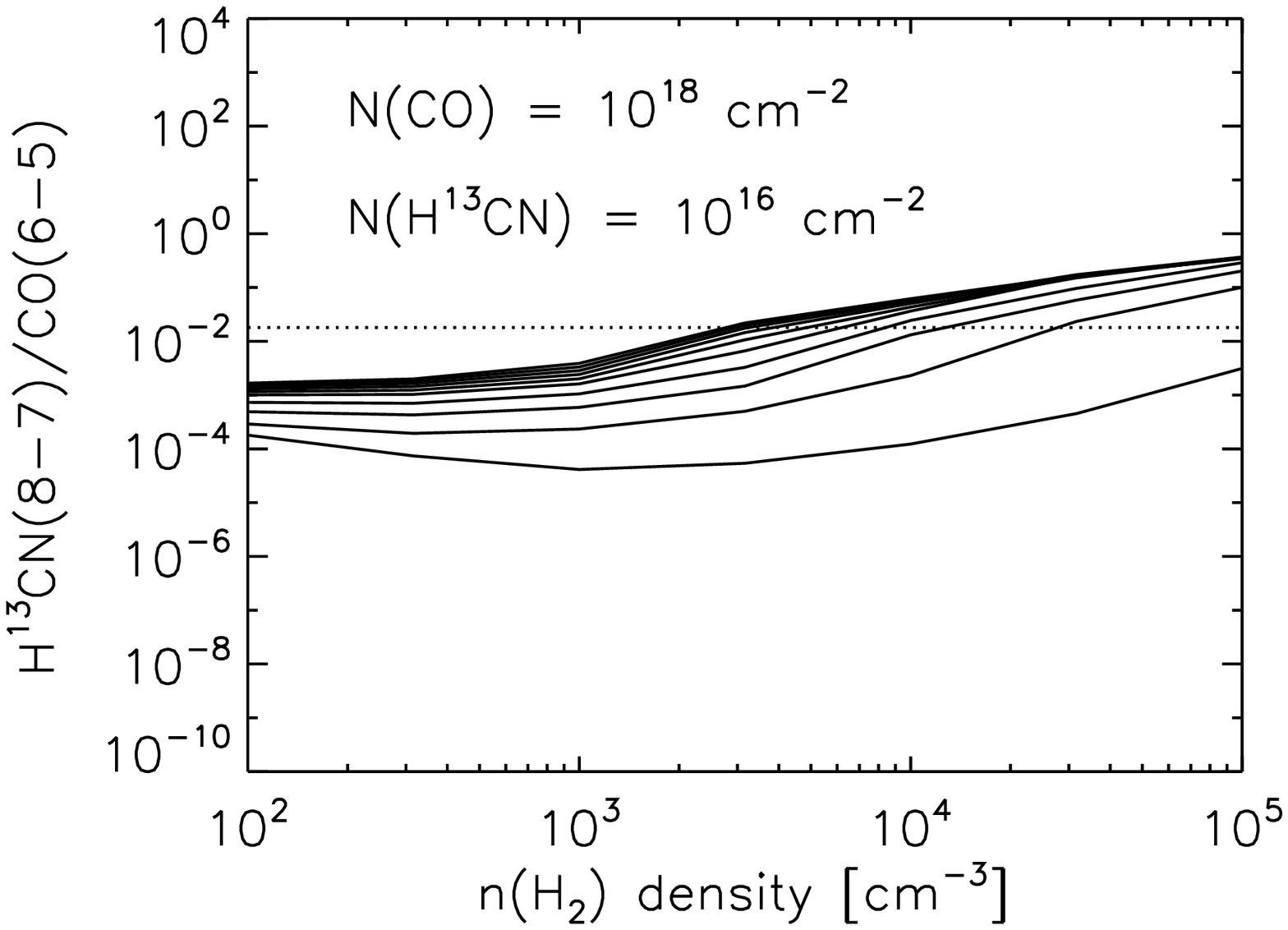}
\caption{Model predictions for the $\rm f_{H^{13}CN(8-7)}/f_{CO(6-5)}$ ratio.
Solid curves present results with different gas temperatures 
of T = 20, 40, ......, 200~K. The dotted line marks the observed
flux ratio between the red-shifted feature and the CO~(6-5) line.
}
\label{fig:model_h13cn}
\end{figure}

The data were reduced and calibrated using CASA 3.4. 
The visibilities were imaged with natural weighting and cleaned.
Finally, phase and amplitude self-calibrations were 
carried out to improve
the overall sensitivity.  Two data sets were generated from the
observations. In the first data set, the CO~(6-5) line data cube was
generated using only data in SPW-0 (sky-freq = 678.039$\pm 1$ GHz),
which encompass the CO~(6-5) line emission with an effective bandpass of
$\rm \sim 800\; km\; s^{-1}$.  The continuum was estimated using data in
the other 3 SPWs. In the second data set, the CO~(6-5) line data
cube was generated using three SPWs: SPW-0, SPW-1 and SPW-3, centered
at 678.039, 679.801 and 676.271 GHz, respectively. The bandwidth of
this CO~(6-5) line data cube is $\rm \sim 2400\; km\; s^{-1}$,
sufficient for the search of high velocity outflows. For the second data
set, the continuum was estimated using only data in SPW-2 (sky-freq =
674.339$\pm 1$ GHz). The first data set is preferred when there is no
significant evidence for outflows 
of $\rm \delta v \gsim 400\; km\; s^{-1}$, because of 
better continuum estimation and subtraction. The continuum subtraction
was carried out using CASA task {\it UVcontsub}.

In order to increase the signal-to-noise ratio, we binned the spectral
cubes into channels with the width of $\rm \delta v = 34\; km\;
s^{-1}$.  The primary beam of the Band-9 observations is $\sim 8''$. For
the configurations used in our observations, the resulting maps have
good quality only in the central region of $\sim 3''$ (limited by
the uv-coverage). Within this
region, channel maps of the CO~(6-5) line emission have 1-$\sigma$
rms noise of $\rm 5.5\; mJy\; beam^{-1}$.  The rms of the integrated
line emission map is 1.2 $\rm Jy\; beam^{-1}\; km\; s^{-1}$, and the
rms of the continuum map is $\rm 1.9\; mJy\; beam^{-1}$. The synthesized beams
of these maps are nearly identical, having the FWHM of $\rm
0\farcs26\times 0\farcs23$ (corresponding to physical scales of
$\rm 106\; pc \times 94\; pc$) and the P.A. of $46^\circ.3$.  The absolute
pointing accuracy of these ALMA observations is on the order of
$0\farcs1$.

\section{Results}
\subsection{CO~(6-5) line emission}\label{sect:CO}
The CO~(6-5) line emission in the central region of NGC~34
  (Fig.~\ref{fig:integrated}) is in a disk that extends to a radius of
  $0\farcs8$ ($\rm \sim 320$ pc), as shown by the
  3-$\sigma$ contour. The disk is well resolved by the ALMA
beam (Fig.~\ref{fig:psf_line}), with a FWHM of $0\farcs5$ (200 pc).  
It fully covers the ``central disk'' of stars (semi-major axis
$a \simeq 0\farcs8$, $\rm P.A.=4^\circ$) which in turn is inside a
larger ($\sim 8''$) ``exponential disk'', both found in the HST V-band
image by \citet{Schweizer2007}. The V-band disk has an axial ratio of
$b/a \simeq 0.72$, corresponding to an inclination angle of 
$i \simeq 44^\circ$. The CO~(6-5) disk looks more face-on 
($b/a \sim 0.9$), though this could be affected significantly by
beam smearing effects. The integrated line emission
has a peak surface density of $\rm 196\pm 29\; Jy\; km\; s^{-1}
beam^{-1}$ (peak brightness temperature = $\rm 36.8\pm 5.5\; K$) 
and a total flux of
$\rm f_{CO~(6-5)} = 1004\pm 151\; Jy\; km\; s^{-1}$. The
uncertainties are dominated by the flux calibration error.  This
  is consistent with the Herschel FTS measurement of the integrated
  CO~(6-5) line emission of NGC~34, which is $\rm 937 \pm 63 Jy\; km\;
  s^{-1}$ (Lu et al., in preparation), indicating that all of the
  warm molecular gas in the galaxy is concentrated in the central disk
  detected by ALMA.

The first-moment map is shown in Fig.~\ref{fig:1stmom}, and
the channel maps are in Fig.~\ref{fig:channel}.
A significant velocity
gradient along the axis of $\rm P.A. = 345^\circ$ can be seen in
these maps, consistent with a slightly inclined rotating disk. 
Fig.~\ref{fig:pvplot} is a P-V plot of the CO~(6-5) line emission
over a stripe of width $\rm =0\farcs25$ and $\rm P.A. = 345^\circ$,
passing through the central core. It shows that the rotation velocity
first rises steadily in the inner part of the disk and then becomes quite
flat at $\rm radius \gsim 0.''5$, again consistent with a resolved rotating 
gas disk that is mostly confined to $\rm radius \lsim 0.''5$.

Panel $a$ of Fig.~\ref{fig:vdis} presents the spectrum of the CO~(6-5) line
emission in the radio heliocentric velocity domain, measured in the channel
maps with an aperture of $\rm radius=1''$. The double-horn shaped spectrum
is again consistent with a rotating disk. The flux-weighted mean,
a measure of the systematic velocity of the kinematic center
of the warm molecular gas disk, is $\rm 5700\; km\; s^{-1}$ with an
measurement error of $\rm \simeq 7\; km\; s^{-1}$ set by the
velocity resolution of the ALMA observations. The 
FWHM of the spectrum is $\rm 406\; km\; s^{-1}$.

A significant emission feature is detected in the red-shifted wing in the
velocity range of $\rm 6000\; km\; s^{-1} \lsim v \lsim 6200\; km\;
s^{-1}$ (Panels $b$ and $d$ of Fig.~\ref{fig:vdis}).
The integrated flux (between $\rm
6000 \leq v \leq 6200\; km\; s^{-1}$) is $\rm
17.7 \pm 2.1\pm 2.7 \; Jy\;km\; s^{-1}$
(the first error is due to rms noise and the second
due to calibration error).
The contour map of this feature is shown in Fig.~\ref{fig:h13cn}.

There are two possibilities for the cause of this feature.
One is $\rm H^{13}CN\; (8-7)$ line emission
(rest-frame frequency = 690.552 GHz), a very high critical 
density isotopic molecular line 
($\rm n_{crit} = 1.8\times 10^{9}\; cm^{-3}$ and 
$\rm T_{ex} = 153\; K$).
In this scenario, the integrated flux of the feature corresponds to
a $\rm f_{H^{13}CN(8-7)}$-to-$\rm f_{CO~(6-5)}$ ratio of $\rm 0.018\pm 0.002$.
And the location of the peak of the feature corresponds to 
$\rm v = 5745\; km\; s^{-1}$ for the $\rm H^{13}CN\; (8-7)$ line,
in very good agreement with
the mean velocity of the CO~(6-5) line emission (see Fig.~\ref{fig:vdis}).
While no detection of $\rm H^{13}CN\; (8-7)$ 
has been repored previously for any extragalactic source,
a few detections of lower J 
$\rm H^{13}CN$ lines in nearby galaxies 
can be found in the literature \citep{Mauersberger1991, 
Nakajima2011, Sakamoto2013}. Of particular interest is 
detection of the $\rm H^{13}CN\; (4-3)$ line,
with a flux of $\rm f_{H^{13}CN(4-3)} = 40.8 \pm 9.4\; Jy\; km\; s^{-1}$,
in the nuclear region of NGC~4418 
\citep{Sakamoto2013}.  Similar to NGC~34, NGC~4418 is also a local LIRG with a 
putative obscured AGN.
Assuming $\rm f_{H^{13}CN(8-7)}/f_{H^{13}CN(4-3)}=f_{HCN(8-7)}/f_{HCN(4-3)}$
($\rm H^{13}CN$ lines and HCN lines have nearly identical
$\rm T_{ex}$ and $\rm n_{crit}$), and the 
ratio $\rm  f_{HCN(8-7)}/f_{HCN(4-3)} = 0.85\pm 0.20$ found in the
Galactic center circumnuclear disk by \citet{Mills2013} 
is applicable to NGC~4418, we can predict that 
$\rm f_{H^{13}CN(8-7)} = 34.7 \pm 11.4\; Jy\; km\; s^{-1}$
for NGC~4418. Dividing this by 
its CO(~(6-5)) flux measured by Herschel FTS,
 $\rm f_{CO(6-5)} = 1068 \pm 116\; Jy\; km\; s^{-1}$ (Lu et al., in
preparation), we obtain a predicted ratio of
$\rm f_{H^{13}CN(8-7)}/f_{CO(6-5)}=0.022\pm 0.007$ for NGC~4418. 
This is in very good agreement with the ratio observed for NGC~34,
lending support for the hypothesis that 
the emission feature in the CO~(6-5) profile of NGC~34
is associated with the $\rm H^{13}CN\; (8-7)$ line.
Indeed, using models of \citet{Meijerink2011}, we found that the
observed flux ratio can be reproduced with the following reasonable
parameters: (1) turbulent line width 
$\rm \delta v_{turb} = 2.7\; km\; s^{-1}$, 
(2) CO column density $\rm N_{CO}=10^{18}\; cm^{-2}$, (3)
temperature $\rm T > 100\; K$, (4) $\rm H_2$ density $\rm n_{H_2} > 10^{3.5}\;
cm^{-2}$, (5) column density ratio $\rm N_{H^{13}CN}/N_{CO} >
10^{-3}$. Some of these model predictions are presented 
in Fig~\ref{fig:model_h13cn}.
High HCN/CO ratios are typical for warm molecular gas.
As shown by \citet{Meijerink2011}, PDR models with strong
mechanical heating can produce both high temperatures and high HCN/CO
ratios, while heating molecular gas to 
high temperatures with X-rays does not produce high HCN/CO ratios.

Another possibility is that the emission feature is associated with
an outflow of warm dense gas with a mean velocity of $\rm 400\; km\; s^{-1}$
and maximum velocity of $\rm 600\; km\; s^{-1}$.
The contour map of the feature (Fig.~\ref{fig:h13cn}) shows an extended
and lopsided morphology. This seems to be consistent with an outflow,
although the blue-shifted component is missing.
\citet{Schweizer2007} found a high speed neutral
gas outflow of $\rm \delta v_{max} = 1050\; km\; s^{-1}$
in the nuclear region of NGC~34, as revealed by
broad blue-shifted $\rm Na\; I \; D$ lines.
In recent JVLA and CARMA observations, \citet{Fernandez2014}
did not detect any outflow in the nuclear region of 
NGC~34 in HI or in CO~(1-0), and they set
an upper limit for the CO mass outflow of $\rm 40\; M_\sun\; yr^{-1}$
(adopting the CO-to-gas-mass conversion factor for ULIRGs).
Assuming an outflow gas with the same $\rm M_{gas}/L_{CO(6-5)}$ ratio
as that for the static gas (see Section~\ref{section:gasdustmass}), 
a radius of the outflow region
of $\rm r=200\; pc$, a mean outflow velocity of
$\rm 400\; km\; s^{-1}$, and adopting 
a ULIRG conversion factor \citep{Downes1998},
we estimate a mass loss rate of
$\rm \dot{M} = 32\; M_\sun\; yr^{-1}$ from the
integrated flux of the feature using the formalism of 
\citet{Rupke2002}. This is consistent
with the upper limit set by \citet{Fernandez2014}.

We cannot rule out either of the two scenarios, and
the feature in the red-shifted wing of CO~(6-5) profile
can be due to the $\rm H^{13}CN$ line
emission or/and an outflow.
Observations of other molecular lines with
high resolution and high sensitivity are
required to resolve this puzzle.

\begin{figure}[!htb]
\epsscale{1.1}
\plotone{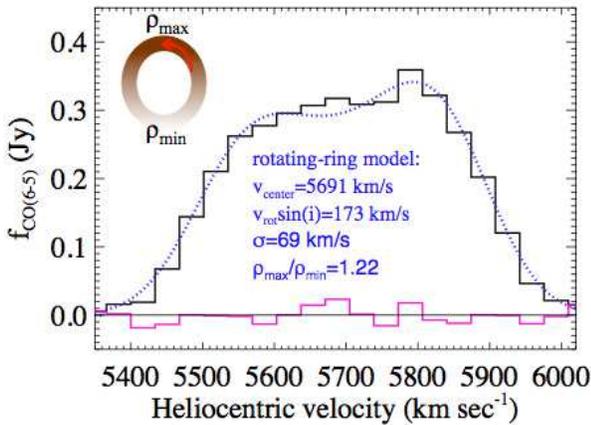}
\caption{
Model fit of the spectrum of the unresolved core. 
The black histogram presents the data and the blue dotted line the model. The
pink histogram near the bottom of the plot presents the residual.
The model assumes a rotating ring with
a central velocity of $\rm v_{center} = 5691\; km\; s^{-1}$,
a projected rotation velocity of
$\rm v_{rot}sin(i) = 173\; km\; s^{-1}$, and a velocity dispersion of
$\rm \sigma = 69\; km\; s^{-1}$. In order to account for 
asymmetric distribution
of the spectrum, the model also assumes 
a density gradient along the major axis of the projected
ring, with the $\rm \rho_{max}/\rho_{min} = 1.22$.
}
\label{fig:vdis_core_fit}
\end{figure}

\begin{figure}[!htb]
\epsscale{1.3}
\plotone{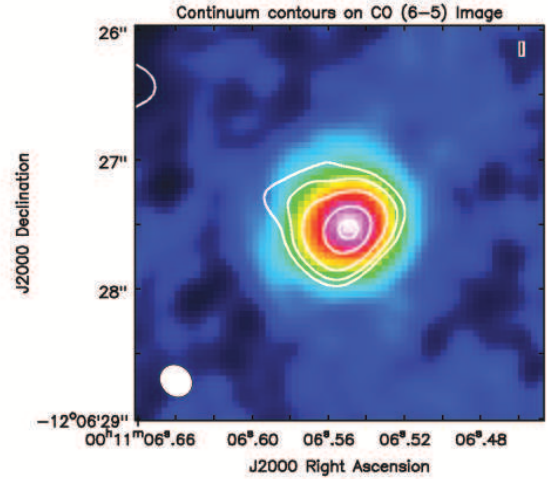}
\vskip-5truecm
\caption{Contours of the dust
continuum at $\rm 435\; \mu m$ overlaid on the CO~(6-5) image. 
The contour levels are 
[1, 2, 4, 8, 12]$\rm \times 5.7\; mJy\; beam^{-1}$.
} 
\label{fig:contimg}
\end{figure}

\begin{figure}[!htb]
\epsscale{1.25}
\plotone{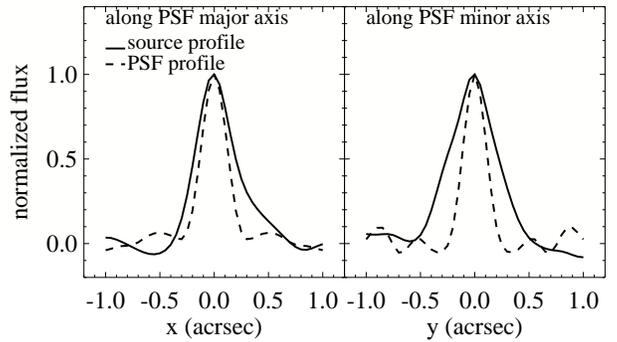}
\caption{Comparisons of profiles of the 435~$\mu m$
dust continuum emission and the synthesized beam.
}
\label{fig:psf_cont}
\end{figure}

The spectrum of the central unresolved core, measured using an
aperture of $\rm r=0\farcs125$, is shown in the Panel $c$ of
Fig.~\ref{fig:vdis}.
Compared to the spectrum of the entire central gas disk,
the mean and the FWHM are nearly identical but the
double-horn shape is less symmetric. The red peak 
is significantly more prominent than the blue peak.  
Different from the large aperture spectrum in Panel $a$,
the spectrum in the core shows a weak blue-shifted wing between 
$\rm 5250 \leq v \leq 5450\; km\; s^{-1}$. 
In Fig.~\ref{fig:vdis_core_fit}, we present model fitting 
to the core spectrum using a simple rotating ring model.
In order to account for the asymmetric distribution
of the spectrum, a density gradient along the major axis of the projected
ring is introduced, with the ratio of $\rm \rho_{max}/\rho_{min}$ being
a free parameter. The least-squares fitting results for
the model parameters are as follows: (1) the
central velocity of the ring: $\rm v_{center} = 5691\pm 18\; km\; s^{-1}$;
(2) the projected rotation velocity of the ring:
$\rm v_{rot}sin(i) = 173\pm 1\; km\; s^{-1}$ (we did not attempt to
constrain the inclination angle $i$ separately); 
(3) the intrinsic (Gaussian) 
velocity dispersion: $\rm \sigma = 69\pm 7\; km\; s^{-1}$; 
(4) the ratio between the maximum and minimum density
in the ring: $\rm \rho_{max}/\rho_{min} = 1.22\pm 0.30 $. 
This simple model gives a reasonably good fit to the data 
including the blue-wing. Other models, including a model with
two separate Gaussians and another model with a rotating ring plus a Gaussian,
can fit the data equally well but with more parameters.

\begin{deluxetable*}{ccccccccccc}
\tabletypesize{\normalsize}
\setlength{\tabcolsep}{0.05in} 
\tablecaption{CO~(6-5) and 435$\mu m$ Continuum in NGC~34 \label{tbl:gas}}
\tablehead{
& (1) & (2) & (3) & (4) & (5) & (6) & (7) & (8) & (9) & (10)\\
\hline
emission & Size & b/a & P.A. & $\rm \delta v$ (FWHM)& $\rm T_{peak}$ & $\rm f_{ALMA}$ & $\rm f_{ALMA}/f_{Hsch}$ & $\rm log(L)$ & $\rm log(M)$ & $\rm log(\Sigma_{peak})$\\
\hline
& (pc) &     & (deg)& ($\rm km\; s^{-1}$)   & (K)           & ($\rm Jy\; km\; s^{-1}$) &     & ($\rm K\; km\; s^{-1}\; pc^2$)  & ($\rm\; M_\sun$) & ($\rm\; M_\sun\; pc^{-2}$) \\
&      &     &      &                      &               & (mJy)                   &     & ($\rm\; L_\sun$) &  &  
}
\startdata
CO~(6-5) & 200 &$\sim 0.9$& 345 & 406 & 36.8$\pm 5.5$ & 1004$\pm 151$ & 1.07$\pm 0.18$ & 8.68$\pm 0.06$
  & $\rm 9.13\pm 0.36$ & $\rm 4.37\pm 0.36$ \\
continuum & 200 &$\sim 0.7$& 315 & & 3.9$\pm 0.6$ & 275$\pm 41$ & 0.53$\pm 0.08$ & 8.63$\pm 0.06$
  & 6.97$\pm 0.13$ & 2.36$\pm 0.13$
\enddata
\tablecomments{
Column (1) -- size (FWHM) of the emission region;
(2) -- minor-to-major axis ratio; (3) -- position angle;
(4) -- FWHM of the velocity distribution; (5) -- peak brightness temperature;
(6) -- for the CO~(6-5): integrated flux in $\rm Jy\; km\; s^{-1}$,
for the continuum: flux density in $\rm mJy$; 
(7) -- ratio between the flux measured
by ALMA (interferometer) and by Herschel (single dish);
(8) -- for the CO~(6-5): logarithm of the CO~(6-5) luminosity in
$\rm K\; km\; s^{-1}\; pc^2$,
for the continuum: logarithm of 
the 435$\mu m$ continuum luminosity in $\rm\; L_\sun$;
(9) -- for the CO~(6-5): logarithm of the
gas mass, which is the geometrical
mean of two estimations based on the Galactic conversion factor 
\citep{Bolatto2013} and the ULIRG conversion factor \citep{Downes1998},
respectively, and the error corresponds to the half difference between
the two; for the continuum: logarithm of the dust mass; 
(10) -- for the CO~(6-5): logarithm of the peak column density of the gas mass,
again an average of the two estimations based on the Galactic conversion factor 
and the ULIRG conversion factor, respectively;
for the continuum: logarithm of the peak dust column density.
}
\end{deluxetable*}

\subsection{Dust continuum emission}
A contour map of the continuum at $\rm 435\; \mu m$ is overlaid on
the CO~(6-5) image
in Fig.~\ref{fig:contimg}.  The centers of the
two maps coincide very well with each other.
The continuum was detected only in the
inner part of the CO~(6-5) disk, and is
slightly elongated with a $\rm P.A. = 315^\circ$.
Both the size and the orientation of the continuum emission region
agree well with those of the ``nuclear bulge'' found in the
HST V-band image by \citet{Schweizer2007}, which is also visible
as a ``cusp'' in the H-band \citep{Haan2013}.
Indeed, there is a hint that the CO~(6-5) line emission may have
a central component corresponding to the same nuclear bulge,
plus a more extended component (undetected in the continuunm) that
corresponds to the V-band nuclear disk.
There is an interesting continuum feature, extendng $\sim 0\farcs7$
 toward the north-east direction from the center, which has
no correspondance in the
CO~(6-5) map, nor in high resolution HST maps. 
Like the line emission, the continuum is also well resolved by
the ALMA beam (Fig.~\ref{fig:psf_cont}),
with a peak surface density of $\rm 76.2\pm 11.4\; mJy\; beam^{-1}$
(peak brightness temperature = $\rm 3.9\pm 0.6\; K$) and a FWHM size of
$0\farcs5$. The total flux density of
the continuum is 
$\rm f_{435\mu m}=275\pm 41\; mJy$. NGC~34 was observed by Herschel-SPIRE 
\citep{Griffin2010} both
in the photometry mode (Chu et al., in preparation) and in
the FTS mode (van~der~Werf et al., in preparation). Because the
error of the continuum measured in the FTS mode is large ($\rm \sim 1\; Jy$),
we estimated the total flux of the 435 $\rm \mu m$ continuum of NGC~34 
by interpolating SPIRE photometer fluxes $\rm f_{350\mu m, SPIRE}=1095\; mJy$
and $\rm f_{500\mu m, SPIRE}=303\; mJy$, and found $\rm f_{435\mu m, SPIRE}=517\; mJy$.
The ratio between $\rm f_{435\mu m}$ and $\rm f_{435\mu m, SPIRE}$ then yields
an interferometer-to-single-dish flux ratio of $\rm 0.53\pm 0.08$. 
This indicates a diffuse component much more extended than the central disk,
contributing up to $\sim 50\%$ of the cold dust emission in NGC~34.

\subsection{Gas Mass and Dust Mass}\label{section:gasdustmass}


\citet{Fernandez2014} recently observed the CO~(1-0) emission of NGC~34 using
CARMA with a synthesized beam of $2\farcs48 \times 2\farcs12$. The integrated 
flux in the central beam of the CO~(1-0) map
is $\rm f_{CO(1-0)} = 42\pm 4\; Jy\; km\; s^{-1}$,
corresponding to a CO~(1-0) luminosity of 
$\rm L_{CO(1-0)} = (7.1\pm 0.7)\times 10^8\; K\; km\; s^{-1}\; pc^{-2}$.
Assuming that this CO~(1-0) luminosity is concentrated in the 
central disk detected by ALMA, which has a physical size 
comparable to that of 
the CARMA beam $\rm (\sim 800\; pc)$,
and adopting a Galactic conversion factor $\rm
\alpha_{CO} = 4.3\; M_\sun\; (K\; km\; s^{-1}\; pc^2)^{-1}$ \citep{Bolatto2013}, 
the gas mass in the central disk is estimated to be
$\rm M_{gas} = 10^{9.49\pm 0.04}\; M_\sun$.
If instead adopting the
conversion factor advocated by \citet{Downes1998} for ULIRGs, $\rm
\alpha_{CO} = 0.8\; M_\sun\; (K\; km\; s^{-1}\; pc^2)^{-1}$, a much
lower gas mass is found: $\rm M_{gas} =10^{8.76\pm 0.04}\; M_\sun$.


The dust mass can be estimated from the continuum flux
using the following formula:
\begin{eqnarray}
\rm M_{dust} & = & \rm {c^2 f_{435\mu m}  D^2\over 2\nu^2kT\kappa_{690GHz}} 
               \nonumber \\
           & = & \rm 9.2\times 10^{6}\; M_\sun\times \nonumber \\
           &   &\rm \left({f_{435\mu m}\over 275\; mJy}\right)
               \left({D\over 84\; Mpc}\right)^2
               \left({T\over 43.0\; K}\right)^{-1}
               \left({\kappa_{690GHz}\over 0.16\; m^2kg^{-1}}\right)^{-1} \nonumber 
\end{eqnarray}
where $\rm \kappa_{690GHz}$ is the mass-opacity at 690 GHz
(central frequency of the continuum), for which we assumed
$\rm \kappa_{690GHz} = \kappa_{1200GHz}\times (690/1200)^2$ and 
$\rm \kappa_{1200GHz} = 0.48\; m^2kg^{-1}$ \citep{Dale2012}.
The dust temperature, $\rm T= 43.0\; K$, is estimated using
the formalism of \citet{Scoville2012} for extremely obscured
nuclear starbursts (i.e. optically thick for the IR emission of 
$\rm \lambda \lsim 100\; \mu m$\footnote{The gas column density profile
shown in Fig.~\ref{fig:mass_profile} confirms that indeed the
central dusty region is optically thick at $\rm \lambda \lsim 100\; \mu m$.}): 
$\rm T_{dust} = 630\times [L_{IR}/10^{12}L_\sun]^{1/4}/R_{pc}^{1/2}\; 
K$. Here we assume the radius of the dusty region
is $\rm R_{pc} = 100$, and 70\% of the IR luminosity of NGC~34 is inside
this radius (as hinted by the high resolution radio continuum data). 
Changing the fraction to 50\% or 100\% will make a difference in
the dust temperature of $\rm \pm 4$ K, corresponding to a 10\% error.
Given the 30\% error for $\rm \kappa$ \citep{James2002},
15\% flux calibration error and $\sim 10\%$ error in the temperature
estimate, the dust mass is $\rm M_{dust} = 10^{6.97\pm 0.13}\; \; M_\sun$.

\section{Comparisons with Previous Results}
NGC~34 is a late stage merger with prominent tidal tails
\citep{Mazzarella1993, Schweizer2007}. 
\citet{Schweizer2007} carried out a detailed analysis of NGC~34
and their results are consistent with
a merger of two gas-rich galaxies with mass ratio of $\rm m/M \sim 1/2$ -- 1/3.
The disks of two galaxies are already coalesced.
An early MIR observation of \citet{Miles1996} found evidence
for a second nucleus about $1''.2$ south of the primary nucleus,
which might coincide with a weak radio source 
in the high resolution VLA radio continuum map \citep{Condon1991}.
However, such a source was not found in the HST NICMOS K-band and
H-band images \citep{Schweizer2007, Haan2011}. In our ALMA
observations, we did not detect this source in the CO~(6-5) line
emission or in the dust continuum, either. Therefore, our observations
are in good agreement with the NICMOS observations and is
consistent with a separation limit of $\rm \lsim 50\; pc$
for any possible double nuclei configuration.

The most solid evidence for the existence of an AGN in NGC~34 is found
in the X-ray \citep{Guainazzi2005, Shu2007, Brightman2011a,
  Esquej2012}.  Recent XMM-Newton observations \citep{Brightman2011a,
  Esquej2012} found an X-ray luminosity of $\rm L_{X,2-10keV} =
1.4^{+0.4}_{0.2}\times 10^{42}\; erg\; s^{-1}$, after the correction
for absorption by a neutral gas column density of $\rm N_{H} =
7.5^{+3.1}_{-2.1}\; 10^{23}\; cm^{-2}$.  Such a high X-ray luminosity
cannot be explained by a pure starburst of SFR $\rm < 200\; M_\sun
yr^{-1}$ \citep{Ranalli2003}.  

Any possible contribution from the AGN to the CO~(6-5)
line emission and the continuum emission
should be inside the unresolved core ($\rm radius = 50\; pc$). 
For the line emission,
the ratio of core flux and the total flux of the central disk
is $\rm f_{CO~(6-5),core}/f_{CO~(6-5)} = 19\%$, and for
the continuum $\rm f_{435\mu m,,core}/f_{435\mu m} = 28\%$.
Hence, even if the AGN would be fully responsible for
the heating of these ``cores'', one could still conclude
that the AGN contribution to total emission
in both cases is insignificant.
This is consistent with previous studies that found low AGN
contribution (in the range of 1 -- 10\%) to the total IR emission of
NGC~34 \citep{Prouton2004, Vega2008, Esquej2012,  Murphy2013, Stierwalt2013}.  
The VLBI
observations of \citet{Lonsdale1993} did not detect any unresolved
core. This sets an upper limit for the core flux of the radio continuum
at 18 cm of $\rm S_{core} < 2.5\; mJy$, suggesting that the AGN
contributes $\lsim 5\%$ of the total radio continuum flux.

\begin{figure}[!htb]
\epsscale{1.25}
\plotone{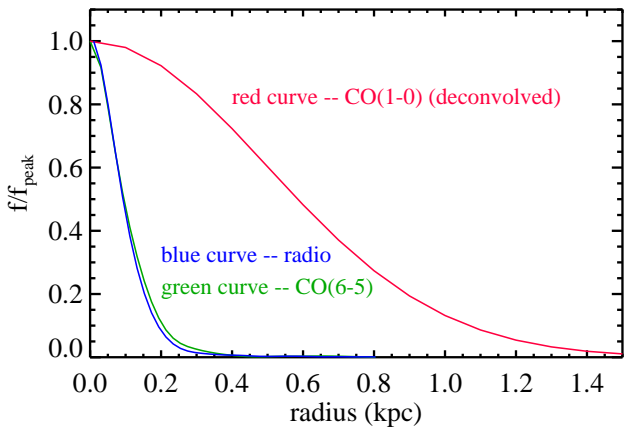}
\caption{Comparisons of normalized radial profiles of the CO~(6-5),
radio continuum  \citep{Condon1991}, and PSF-deconvolved CO~(1-0)
\citep{Fernandez2014}.
}
\label{fig:comp_co65_10_radio}
\end{figure}


The star formation in  NGC~34 is concentrated in a nuclear
starburst, which dominates the IR and radio luminosities of
the LIRG \citep{Condon1991, Miles1996, Gorjian2004, Esquej2012}.
\citet{Esquej2012} estimated the size of the starburst region
is $\sim 0.5$ -- 1~kpc (corresponding to
1\farcs25 -- 2\farcs5 FWHM), and did not find any significant
IR emission beyond 2~kpc (5\arcsec FWHM). Our higher
angular resolution ALMA maps revealed a central disk
of FWHM = 0\farcs5 (0.2 kpc), significantly smaller
than the MIR emitting region studied by 
\citet{Miles1996} and \citet{Esquej2012}.
In Fig~\ref{fig:comp_co65_10_radio} we compared the
radial profiles of the CO~(6-5) and radio continuum
at 8.44 GHz  ($\rm beam = 0\farcs39\times 0\farcs29$, \citealt{Condon1991})
and found very good agreement\footnote{There is a position offset
of $\sim 0\farcs1$ between peak of the CO~(6-5) and
that of the radio continuum, which we attributed to the
astrometrical error of the ALMA observations.}.  
Since the AGN contribution to the
radio continuum is negligible \citep{Lonsdale1993}, this suggests
a tight correlation between the warm gas probed by the CO~(6-5) 
and current star formation activity probed by the 
radio continuum on the time scale of $\sim 10^8\; yrs$ \citep{Xu1990}.
It is unlikely that any substantial
fraction of the radio continuum is missing in the high resolution 8.44 GHz map 
because the total flux obtained using the map fits very well
in the spectral index analyses \citep{Clemens2008, Fernandez2014}, 
together with radio fluxes in other bands of much lower resolutions (e.g. 
$\rm f_{1.4GHz}$ with beam size of 45\arcsec, \citealt{Condon1998}).
The larger size of the MIR emission could be due to contributions
of older stars (e.g. those in the central disk of 
$\rm \sim 4\times 10^{8}  yrs$ old; \citealt{Schweizer2007}) to
the PAH emission \citep{Lu2003}.

The CO~(6-5) disk is also significantly more compact than the CO~(1-0)
disk found by \citet{Fernandez2014}: after deconvolving with a
Gaussian beam of $\rm 2\farcs48\times 2\farcs14$
\citep{Fernandez2014}, the CO~(1-0) emission has a half-peak radius of 0.6~kpc
which is about 6 times of that of the CO~(6-5) emission
(Fig~\ref{fig:comp_co65_10_radio}). It appears that a large fraction of
the CO~(1-0) emission is not associated with  
current star formation activity (as indicated by the radio continuum).
This is consistent with  previous 
studies which suggested that the low J CO luminosities may be dominated by the
emission of diffuse gas not closely related to the active star
formation regions \citep{Scoville1997, Downes1998, Bryant1999,
Gao2001b}. The SMA CO~(3-2) observations 
have also shown more compact emission regions
\citep{Wilson2008, Ueda2012, Sakamoto2013} and closer correlation with
the star formation activity compared to the CO~(1-0) emission
\citep{Yao2003, Iono2004, WangJ2004, Iono2009}.
Probing even higher density gas than
the CO~(3-2) line does, the CO~(6-5) emission has perhaps more 
in common with dense molecular lines such as HCN~(1-0) 
($\rm n_{crit} = 2.6\times 10^{6}$), for which \citet{Gao2004a}
found a very tight and linear correlation with the SFR.

\begin{figure}[!htb]
\epsscale{1.25}
\plotone{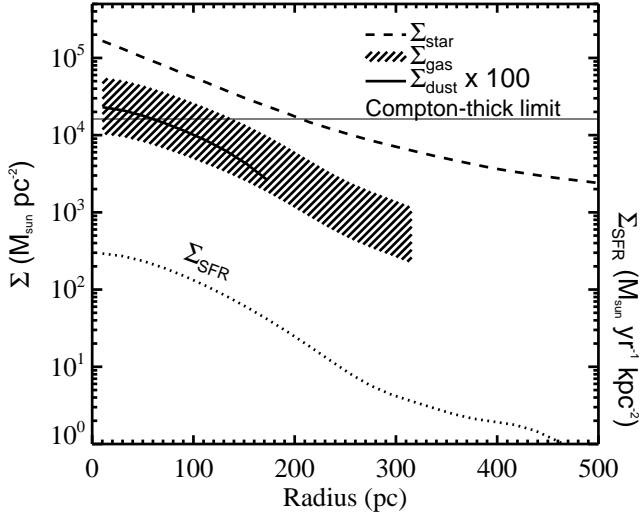}
\caption{Radial profiles of column densities of stars, gas,
  and dust (multiplied by a factor of 100).  
  The stellar mass is estimated using
  the HST NICMOS H-band data \citep{Haan2011}.  The upper boundary of
  the stripe representing the gas mass column density profile is estimated
  from ALMA CO~(6-5) observations using the Galactic conversion
  factor \citep{Bolatto2013} and the lower boundary the ULIRG
  conversion factor \citep{Downes1998}. The dust mass column density is
  derived from ALMA 435$\mu m$ continuum continuum observations.
  The Compton-thick limit corresponds to a gas column density of 
  $\rm N_H = 1.5\times 10^{24}\; cm^{-2}$. 
}
\label{fig:mass_profile}
\end{figure}

In Fig.~\ref{fig:mass_profile} the radial profiles of
the gas and dust column densities are presented. These were 
estimated using radial profiles of the CO~(6-5) and the continuum and
assuming constant mass-to-luminosity ratios.
Here we adopted, according to values listed in
 Table~\ref{tbl:gas},  a $\rm M_{gas}/L_{CO(6-5)}$ ratio of 5.3 (1.0) 
$\rm M_\sun\; (K\; km\; s^{-1}\; pc^2)^{-1}$ for Galactic (ULIRG) conversion
factor and a $\rm M_{dust}/L_{435\mu m}$ ratio of $\rm 0.022\; M_\sun\; L_\sun^{-1}$.
The plot shows very
high gas column density in the inner most region of $\rm r< 100\; pc$:
the peak $\rm \Sigma_{gas}$ derived using the Galactic conversion factor
is about 2.5 times above the Compton-thick limit 
($\rm N_H = 1.5\times 10^{24}\; cm^{-2}$), and that
derived using the ULIRG conversion factor is only a factor of 2 lower
than the Compton-thick limit.
It is worth noting that the central gas column density estimated
using ALMA data is consistent with that estimated using the X-ray data
\citep{Brightman2011a, Esquej2012}. 

For comparison, we also plot in Fig.~\ref{fig:mass_profile} the $\rm
\Sigma_{SFR}$ profile estimated using the radio continuum at 8.44 GHz and
the $\rm \Sigma_{star}$ profile derived using the HST NICMOS H-band
data (angular resolution: $0.''15$; \citealt{Haan2011}).  In order to
convert $\rm P_{8.44GHz}$ to the SFR, we assumed that $\rm
S_{1.4GHz}/S_{8.44GHz}=3.86$ \citep{Condon1991}, $\rm q =
log(L_{fir}/(3.75\times 10^{12}\; Hz)/P_{1.4GHz}) = 2.34$
\citep{Yun2001}, $\rm L_{IR}(8$ -- $\rm
1000\mu m)/L_{fir}(40$ -- $\rm 120\mu m) = 1.8$, and $\rm SFR/(M_\sun\; yr^{-1}) =
L_{IR}/(1.16\times 10^{10}\; L_\sun)$ (\citealt{Kennicutt1998a};
converted to a Kroupa IMF).  For the estimate of 
$\rm \Sigma_{star}$, we assumed that
the stellar mass in the inner most $\sim 500\;
pc$ of NGC~34 is dominated by the starburst, 
as suggested by the simulation results of
\citet{Hopkins2008b}, and adopted a constant H-band
luminosity-to-mass ratio predicted by a single-burst stellar
population (SSP) model of \citet{Bruzual2003} with the Kroupa IMF,
solar metallicity, and starburst age of $\rm \tau = 4\times 10^8\; yr$
\citep{Schweizer2007}. An H-band extinction correction of $\rm
A_{H}=0.46$ mag was applied, derived from the $\rm Pa\beta$-to-$\rm
Br\gamma$ ratio \citep{Rodriguez2005} and the extinction curve of
\citet{Calzetti2000}\footnote{The $\rm \Sigma_{dust}$ plotted in
  Fig.~\ref{fig:mass_profile}, if uniformly distributed in a
  foreground screen, indicates much higher $\rm A_{H}$. However, it is
  likely that the dust is clumpy and much of the nuclear bulge is in
  front of dust.}. Interestingly, the star-formation
timescale derived from $\rm \Sigma_{star}/\Sigma_{SFR}$ ratio is
rather constant in the inner part of the disk of $\rm r \lsim 200\;
pc$, with a value of $\rm \sim 8\times 10^8\; yr$. Given the
uncertainty of $\rm \Sigma_{star}$, which is on the order of a factor of 2
(due to errors in both the assumed light-to-mass ratio and the extinction
correction),
this is marginally higher than the
starburst age ($\rm \tau = 4\times 10^8\; yr$) 
found by \citet{Schweizer2007} using the optical colors, consistent with
a relatively old starburst already
passed its peak activity. Indeed, the
comparison between $\rm \Sigma_{SFR}$ and $\rm \Sigma_{gas}$ yields a
gas exhaustion time scale of $\rm M_{gas}/SFR \sim 1.6\; (0.3) \times 10^8\; yr$ 
if the Galactic (ULIRG) conversion factor is used, which is
significantly shorter than the star formation time
scale.  Nevertheless, the starburst is still very strong.  With $\rm
\Sigma_{SFR}\simeq 300\; M_\sun\; yr^{-1}\; kpc^{-2}$ and $\rm
\Sigma_{gas}\simeq 2\times 10^{4}\; M_\sun\; pc^{-2}$ at the central peak, it
falls right on the Kennicutt-Schimdt law and
is among the extreme starbursts with the highest $\rm \Sigma_{SFR}$
and $\rm \Sigma_{gas}$ (see Fig.~6 of \citealt{Kennicutt1998b}; but
also see \citealt{Liu2012}).

\section{Summary}\label{sect:summary}
The CO~(6-5) line emission in the central region of NGC~34 
($\rm radius \lsim 500\; pc$) is well
resolved by the ALMA beam ($\rm 0\farcs26\times 0\farcs23$).
This is the first time the warm and dense gas in
the nuclear region of a LIRG is resolved 
with a linear resolution of $\sim 100$ pc.
Both the morphology and kinematics of the CO~(6-5) line
emission are rather regular, consistent with a compact rotating
disk which has a FWHM size of 200~pc and extends to a radius of
$\rm r = 320\; pc$.
The integrated CO~(6-5) line flux of the disk is 
$\rm f_{CO~(6-5)} = 1004 (\pm 151) \; Jy\; km\;
s^{-1}$, recovering all the single dish CO~(6-5) line flux of
NGC~34 measured by Herschel. The line profile shows a double-horn
shape, with a FWHM of $\rm 406\; km\; s^{-1}$.

A significant emission feature is detected on the red-shifted 
wing of the profile,
coincident with the frequency of the $\rm H^{13}CN\; (8-7)$ line
emission (rest-frame frequency = 690.552 GHz), with an
integrated flux of $\rm 17.7 \pm
2.1 (random) \pm 2.7 (sysmatic)\; Jy\;km\; s^{-1}$. However,
it cannot be ruled out that the feature is due to an outflow of warm dense
gas with a mean velocity of $\rm 400\; km\; s^{-1}$.

The $\rm 435\; \mu m$ 
continuum emission is resolved into an elongated configuration
($\rm P.A. =315^\circ$). The
flux is $\rm f_{435\mu m} = 275\; (\pm 41) \; mJy$ ($53\pm 8 \%$ of
the total 435$\mu m$ flux derived from Herschel observations),
corresponding to a dust mass of $\rm M_{dust} = 10^{6.97\pm 0.13}\; M_\sun$.  
An unresolved central core ($\rm radius \simeq 50\; pc$)
contributes $28\%$ of the continuum flux
and $19\%$ of the CO~(6-5) flux detected by ALMA, consistent with
insignificant contributions of the AGN to both emissions.

The CO~(6-5) disk is a factor of 6 smaller than the CO~(1-0) disk 
found by \citet{Fernandez2014}. 
Comparison with radio continuum
suggests that the nuclear starburst has about the same distribution
of the warm and dense gas probed by CO~(6-5), while
much of the diffuse gas probed by CO~(1-0) is not associated with 
star formation. Both the CO~(6-5) line and continuum
distributions indicate a very high gas column density ($\rm \gsim 10^4\;
M_\sun\; pc^{-2}$) in the nuclear region ($\rm radius \lsim 100\; pc$),
consistent with the extremely high SFR density found in
the same region.

\vskip1truecm
\noindent{\it Acknowledgments}:
CKX acknowledges useful discussions with George Privon, Eckard Sturm,
Francois Schweizer, and Ximena Fernandez.  Kim Scott and Tony Remijan
from NAASC are thanked for their helps on data reduction.  Y.G. is
partially supported by NSFC-11173059, NSFC-11390373, and
CAS-XDB09000000. Y.Z. thanks the NSF of Jiangsu Province for partial support under grant
BK2011888. V.C. would like to acknowledge partial support from
the EU FP7 Grant PIRSES-GA-2012-316788. This paper makes use of the
following ALMA data: ADS/JAO.ALMA-2011.0.00182.S. ALMA is a
partnership of ESO (representing its member states), NSF (USA) and
NINS (Japan), together with NRC (Canada) and NSC and ASIAA (Taiwan),
in cooperation with the Republic of Chile. The Joint ALMA Observatory
is operated by ESO, AUI/NRAO and NAOJ.  This research has made
extensive use of the NASA/IPAC Extragalactic Database (NED) which is
operated by the Jet Propulsion Laboratory, California Institute of
Technology, under contract with the National Aeronautics and Space
Administration.

\bibliographystyle{apj}
\bibliography{/Volumes/Seagate/data1/bibliography/papers_biblio}


\end{document}